\documentstyle[12pt,epsf]{article}
\newcommand{\ds }{\displaystyle}
\newcommand{\ra}{\rightarrow}
\newcommand{\be}{\begin{equation}}
\newcommand{\ee}{\end{equation}}
\newcommand{\bea}{\begin{eqnarray}}
\newcommand{\eea}{\end{eqnarray}}
\newcommand{\ci}{\cite}
\newcommand{\bi}{\bibitem}
\newcommand{\nono}{\nonumber \\}

\newcommand{\quart}{\frac{1}{4}}

\newcommand{\dd}{\partial}

\newcommand{{\bfna}}{\mbox{\boldmath$\vec{\nabla}$}}

\newcommand{\half}{\frac{1}{2}}
\newcommand{\al}{\lambda}

\def\dal{\,\lower0.3ex\vbox{\hrule\hbox{\vrule\kern2pt\vbox{\kern4pt\kern4pt}
\kern2pt\vrule}\hrule}\,}

\begin{document}

\title{\sl The Sine-Gordon Wobble}
\vspace{1 true cm}
\author{G. K\"albermann$^*$
\\Soil and Water dept., Faculty of
Agriculture, Rehovot 76100, Israel}
\maketitle

\vspace{3 true cm}
\begin{abstract}

Nonperturbative, oscillatory, winding number one solutions 
of the Sine-Gordon equation are presented and studied numerically. 
We call these nonperturbative shape modes {\sl wobble} solitons.
Perturbed Sine-Gordon kinks are found to decay to {\sl wobble} solitons.

\end{abstract}

$^*${\sl e-mail address: hope@vms.huji.ac.il}

\newpage

\section{\sl Introduction}

The Sine-Gordon equation was discovered in the study of constant 
negative curvature metric spaces at the end of the 19$^{th}$ 
century\ci{eisenhart}. It reappeared in physical problems dealing with
one dimensional dislocations
\ci{lamb}, long Josephson junctions and in other settings.\ci{rem}

The Sine-Gordon equation possesses solitary wave solutions.
These solitary waves are solitons\ci{zabusky}.
The equation is completely integrable and has an infinite number of 
conserved currents.\ci{lamb, rajaraman}

The solitons of the Sine-Gordon theory carry a topological winding number 
{\sl q}. Sine-Gordon solitary waves are topological solitons. 
The winding number zero sector {\sl q=0} 
supports bound soliton-antisoliton
solutions, the breathers, as well as an unbound soliton-antisoliton 
pairs\ci{lamb}. 
The {\sl q=1} sector solitary wave is the kink soliton.

In the recent past, a controversy has arisen concerning the existence 
of oscillatory solutions in the q =1 sector. 
Shape modes were predicted by
Rice\ci{rice} by means of a collective coordinate method. 
Boesch and 
Willis\ci{boesch} studied the excitation of this internal quasimode using
a more refined collective coordinate approach as well as numerical integration.
The predictions of both works are not exact, 
or even approximately so, due to the limitations of the collective coordinate 
method.
In the numerical and collective coordinate treatments, the angular 
frequency of the oscillations of the kink
soliton width was found to be above the threshold for the production
of phonons. Phonons are the solutions of the Klein-Gordon equation, obtained 
by linearizing the Sine-Gordon equation\ci{fogel}.
Hence, if this oscillation exists, it is embedded in the continuum
and must decay, albeit with a small decay constant for angular frequencies
near threshold.

Quintero et al.\ci{quintero} have recently contested the existence
of this shape mode. They have argued that the numerical solution 
of Boesch and Willis\ci{boesch} is incorrect. 
Quintero et al.\ci{quintero} suggest that
this quasi-mode is nothing more than a numerical effect due to discretization.
In discrete nonlinear equations, a mode in the continuum, can sink below 
threshold depending on the value of lattice constant\ci{kivshar}.
\\
\indent
In the present work we show analytical nonperturbative solutions to the
Sine-Gordon equation that are oscillatory and apparently stable that we call 
{\sl the wobble} solitons. The wobble solitons are derived by means of the
Inverse Scattering Transform (IST) method following 
Lamb\ci{lamb} and Segur\ci{segur}.
The IST method produces
soliton solutions based on scattering data of 
Schr\"odinger-like equations. The data leads to a potential - hence the name
inverse scattering transform - from which the soliton is derived. 
The expression for the wobble soliton we derive 
corrects the one given by Segur\ci{segur}.
The solution will be depicted and checked analytically as well as numerically.

The angular frequency of the oscillation of the wobble is found to
range between zero and one, where the phonon continuum takes over.
There is no gap in the frequency spectrum. A dense set of nonperturbative
nonlinear wobbles fills it. We also touch upon the stability issue.

Having shown the existence and probable stability of the wobble, we 
connect to the problem of shape oscillations in distorted
kinks. We recover the results of Boesch and Willis\ci{boesch} and  
point out a probable source 
of error in the numerical calculations of Quintero et al.\ci{quintero}
The shape modes found in the literature are shown to be
an intermediate stage on the way between distorted kinks 
and wobbles.

The next section summarizes the IST derivation of the 
the wobble\ci{lamb},\ci{segur} and addresses the stability issue.
Section 3 deals with the distorted kink problem and the controversy 
around the existence of a kink shape mode in the Sine-Gordon equation.
Conclusions are presented in section 4.

\section{\sl The wobble by the inverse scattering method}

The powerful technique of the Inverse Scattering Transform\ci{lamb},
connects between nonlinear equations, such as the Korteweg-deVries (KdV), 
Sine-Gordon, nonlinear Schr\"odinger, modified KdV, etc., and linear
eigenvalue equations, such as the Schr\"odinger or Dirac-like two-component
equations.
The nonlinear equations arise as consistency conditions on the linear
equations. The potentials of the linear equations yield solutions
of the nonlinear equation. 
The method uses the scattering data
of the linear problem to predict the nonlinear solution by resorting to
an integral equation discovered by Gelfand, Levitan and Marchenko\ci{ablowitz}.

In the case of reflectionless potentials, for which there are only
transmitted waves in the linear problem, 
the Gelfand-Levitan-Marchenko equations
are integrable in closed form. The analytical formulae are given by Lamb
\ci{lamb}. Segur\ci{segur} implemented these formulae for the case
of what we presently call {\sl the wobble}.
The Sine-Gordon soliton composed of a kink and a breather, 
the {\sl wobble} $u(x,t)$ with its center at rest, is given by

\be\label{wobble}
u(x,t)=4~Im(ln(det(I+iM)),
\ee

{\noindent}where $\ds I_{i,j}, i,j=1,3$ is the unit matrix , and $\ds M_{i,j}$
 is the matrix containing scattering data of the kink and the 
breather.

\bea\label{M}
M_{j,k}&=&\frac{-i~m_k}{\zeta_j+\zeta_k}~e^{\theta},\nono
\theta&=&-i~(\zeta_j+\zeta_k)\frac{x+t}{2}+i~\frac{x-t}{4~\zeta_k},
\eea
\noindent
where $\ds \zeta_1=i/2,~\zeta_2=\alpha+i\beta, \zeta_3=-\alpha+i\beta$, with
$\ds \alpha^2+\beta^2=0.25$ and $\ds \beta~>~0$ are the scattering amplitudes:
$\ds \zeta_1$ for the
{\sl q=1} kink and $\ds \zeta_{2,3}$ for the {\sl q=0} breather. $\ds m_j$ are 
the normalization constants for each matrix element, with $\ds m_1$ real and 
$\ds m_2^*=m_3$.
The wobble depends on the parameters 
$\ds m_2=(m_{2R},m_{2I}),~m_1,~\beta$.
A moving soliton can be obtained by boosting 
with a Lorentz transformation. We here focus on a {\sl wobble} at rest.
\\
{\indent}Defining the complex function 
\bea\label{F}
F=det(I+iM)=V+iW,
\eea
\noindent

{\noindent} we find
\footnote{Eq.(\ref{UW}) corrects the results of Segur\ci{segur}}

\bea\label{UW}
V(x,t)&=&1+\frac{|m_2|^2~(\quart-\beta^2)}{\beta^2}e^{4\beta x}-
\frac{2~m_1|m_2|(\half-\beta)}{\half+\beta}~e^{x+2\beta x}~cos(2\alpha (t+t_0))
,\nono
W(x,t)&=&-\frac{m_1~|m_2|^2~(\half-\beta)^3}{\beta^2(\half+\beta)}
e^{x+4\beta x}-~m_1~e^x+2~|m_2|~e^{2\beta x}~cos(2\alpha (t+t_0)),
\eea
{\noindent}where 

\bea\label{cos}
\frac{2}{|m_2|} ~tan(2\alpha t_0)=\frac{m_{2I}\beta+m_{2R}\alpha}
{m_{2I}\alpha-m_{2R}\beta}.
\eea

The {\sl wobble} is  obtained by inserting eq.(\ref{UW}) 
in eq.(\ref{wobble}).
\\
{\indent}As shown below, the {\sl wobble}
oscillates sweeping over values above $\ds 2\pi$.
In numerical codes that limit the inverse tangent to the principal
branch, it is imperative to use the 
complex natural logarithm expression of eq.(\ref{wobble}), 
instead of the translation $\sl Imag(ln(F))=tan^{-1}(\frac{W}{V})$.

The {\sl q=1} kink is recovered from the wobble of 
eqs.(\ref{wobble},\ref{UW}) by setting $\ds m_2=(0,0)$.
The {\sl q=0} breather requires the substitution $m_1=0$.

The normalization constant $\ds m_1$ determines the location of the center
of the wobble, $\ds m_2$ fixes the amplitude of the oscillation and the
phase. 
The angular frequency of the oscillation is $\omega=2~\alpha$, with
upper bound $\omega_{max}=1$, at which the the phonon 
spectrum begins\ci{rubinstein},\ci{fogel}.

An alternative simpler form of $F= V+ i~W$ is
 
\bea\label{UW1}
V(x,t)&=&2~e^{2\beta x}\bigg(e^{\mu}cosh(2\beta x+\mu)-|m_2|
e^{x+\al-\mu}~cos(2\alpha (t+t_0))\bigg),\nono
W(x,t)&=&2~e^{2\beta x}\bigg(-m_1~e^{x+\al}~cosh(2 \beta x+\al)
+|m_2|~cos(2\alpha (t+t_0))\bigg),
\eea
\noindent
where $\ds 
e^{\al}=\frac{|m_2|}{|\beta|}~\sqrt{\frac{(\half-\beta)^3}{\half+\beta}}$,
 $\ds e^{\mu}=\frac{|m_2|}{|\beta|}\sqrt{\quart-\beta^2}$.

It is an arduous but straightforward task to show that the wobble obeys the
renormalized Sine-Gordon equation.\footnote{We have verified 
eq.(\ref{UW1}) using computerized algebra.}

\be\label{sg}
\frac{\dd^2 u}{\dd t^2}-\frac{\dd^2 u}{\dd x^2} + sin(u) = 0,
\ee

derivable from the renormalized lagrangian

\bea\label{lag}
{\cal L}=\int{~dx~\bigg[\bigg(\frac{\dd u}{\dd t}\bigg)^2
-\bigg(\frac{\dd u}{\dd x}\bigg)^2+(cos(u)-1)}\bigg],
\eea
\noindent
with renormalized energy 

\bea\label{energy}
E~=~8~+32\beta.
\eea

\begin{figure}
\epsffile{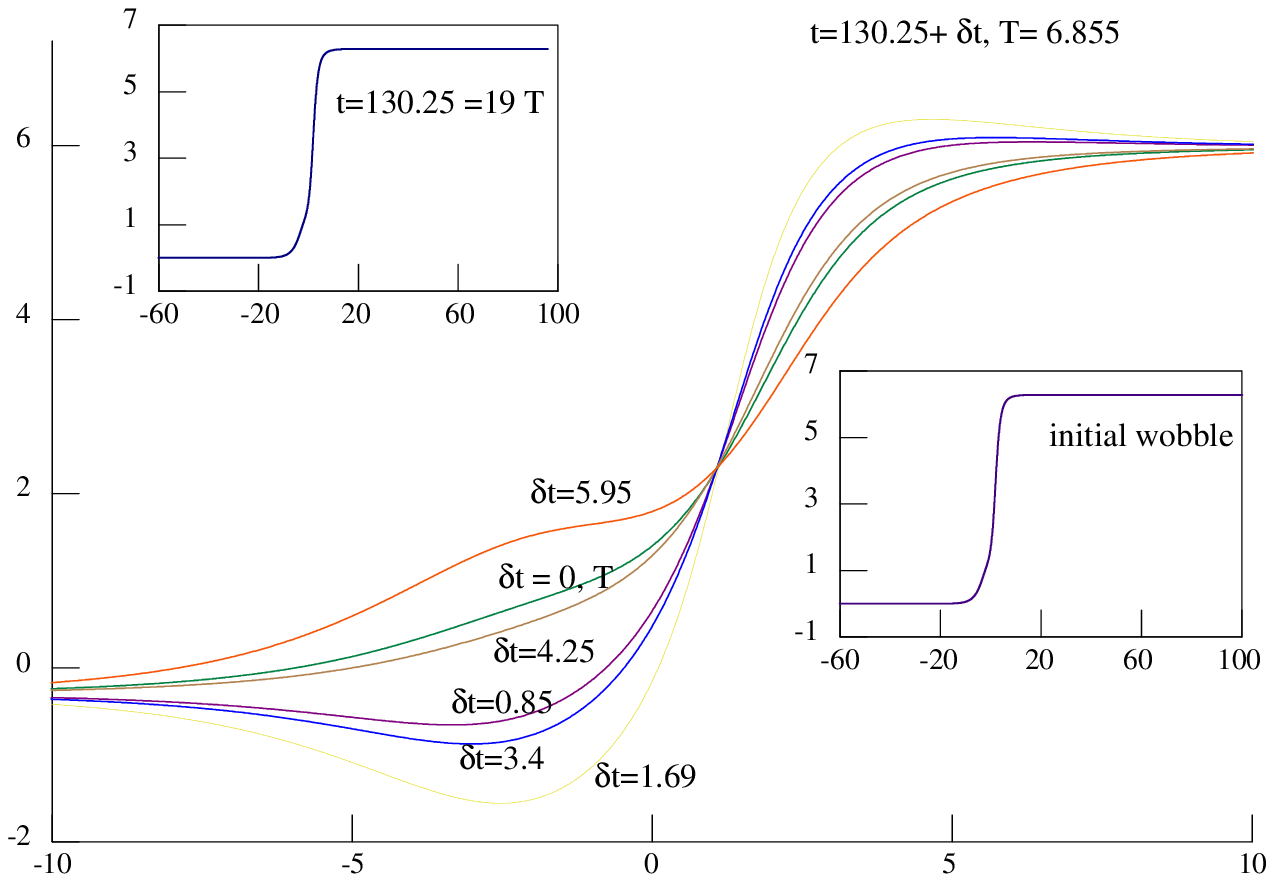}
\caption{\sl Wobble as a function of distance: $m_1=-1,
 m_2=(0.6,0.7),\beta=0.2$} 
\label{fig1}
\end{figure}
The remarkable IST method has yielded a new nonlinear 
solution of the Sine-Gordon equation built from a {\sl q=1}
kink and a {\sl q=0} breather, a nonperturbative oscillatory shape mode of
the kink.

\begin{figure}
\epsffile{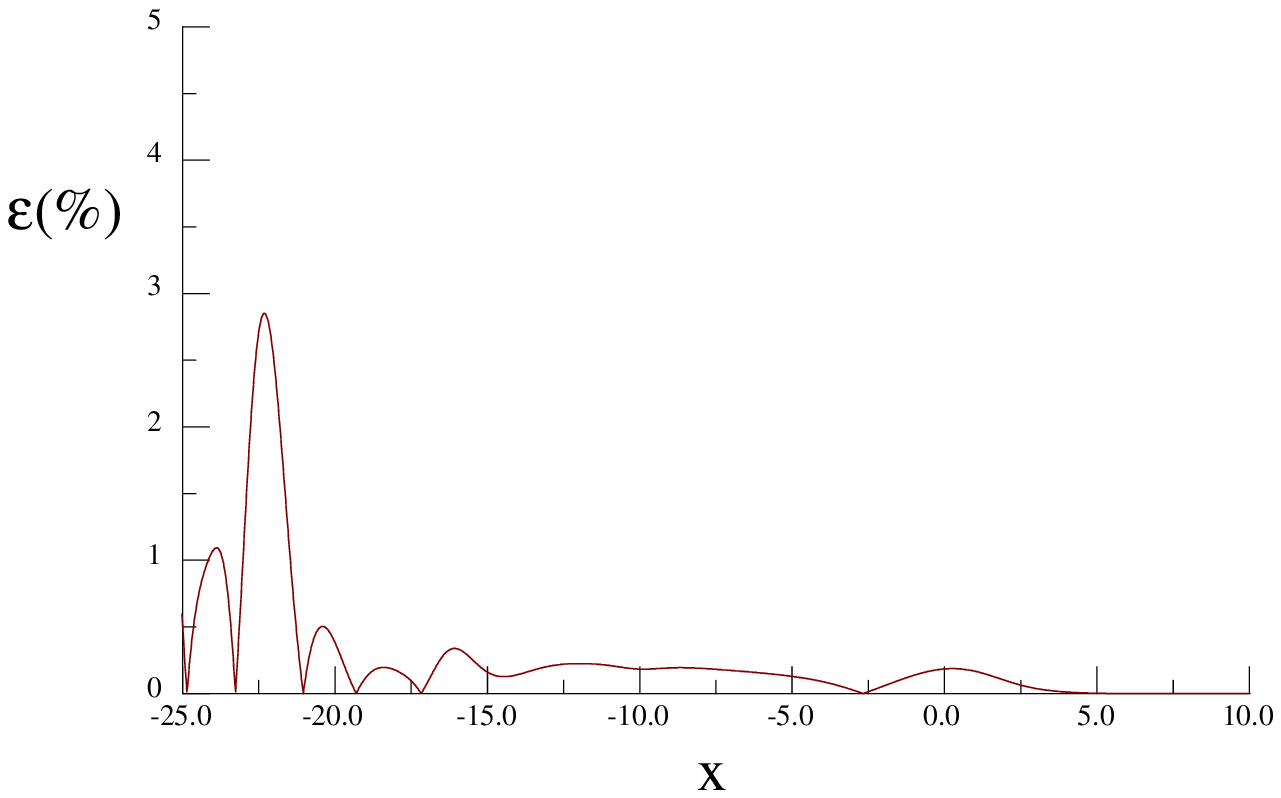}
\caption{\sl $\epsilon$ as a function of distance for the wobble parameters 
of figure 1}
\label{fig2}
\end{figure}

Figure 1 shows snapshots of the wobble at 
times comprising one oscillation period. 
As eq.(\ref{UW1}) implies, the curves show  
oscillations that are not simple harmonic.
The insets in figure 1
show the wobble at t=0 and and an identical picture after 19 periods.
\\
The numerical integration code we used is based
on the leapfrog method. The calculation used double precision variables with
a fixed time step of dt = 0.01 and a flexible space grid.
One measure of the integration accuracy is the
conservation of the energy. The error in the energy was demanded 
never to exceed 0.2\%. The other measure consisted in an exact match 
with the analytical formulae of eq.(\ref{UW1}) for long times.
Figure 2 shows one such a comparison for the wobble parameters
of figure 1 at t=137 amounting to 13700 iterations of the numerical code. 
The ordinate is the absolute value of the percentage 
relative deviation of the numerical
results $\ds u(x,t)_{num}$ from the analytical
formula of eq.(\ref{wobble}) $\ds~u(x,t)_{form}$, 
$\ds~\epsilon 
=100~abs\bigg(\frac{u(x,t)_{num}-u(x,t)_{form}}{u(x,t)_{form}}\bigg)$.
The abscissa spans the region where the results are relevant. Below {\sl x=25}
the wobble is negligible and the comparison is irrelevant. The errors very 
rarely exceed one percent. 
\\
{\indent}The stability problem of distorted solitons under large perturbations 
has not been settled yet, even for the case of the breather\ci{birnir}.
The study of wobble stability can be circumscribed to
the analysis of its development for $\ds \alpha,\beta$ 
violating the unitarity condition $\ds \alpha^2+\beta^2=\quart$. 
These parameters belong to the breather sector of the wobble.

Initially we addressed the stability problem by using a limited set
of collective coordinates, promoting the parameters to be time dependent.
Unfortunately, the method failed to predict the observed behavior.
Even the frequency of the unperturbed breather or wobble cannot be recovered
by means of the collective coordinate approach. 

We therefore 
proceeded to investigate the question of stability of both the breather and 
the wobble by means of numerical simulations.
We scanned the parameter ranges $\ds -\infty<\alpha
<\infty $, $\ds -0.5<\beta<0.5$, omitting the $|\beta| > 0.5$ 
region, for which the distorted
breather and wobble decay by emission of soliton-antisoliton pairs.

\begin{figure}
\epsffile{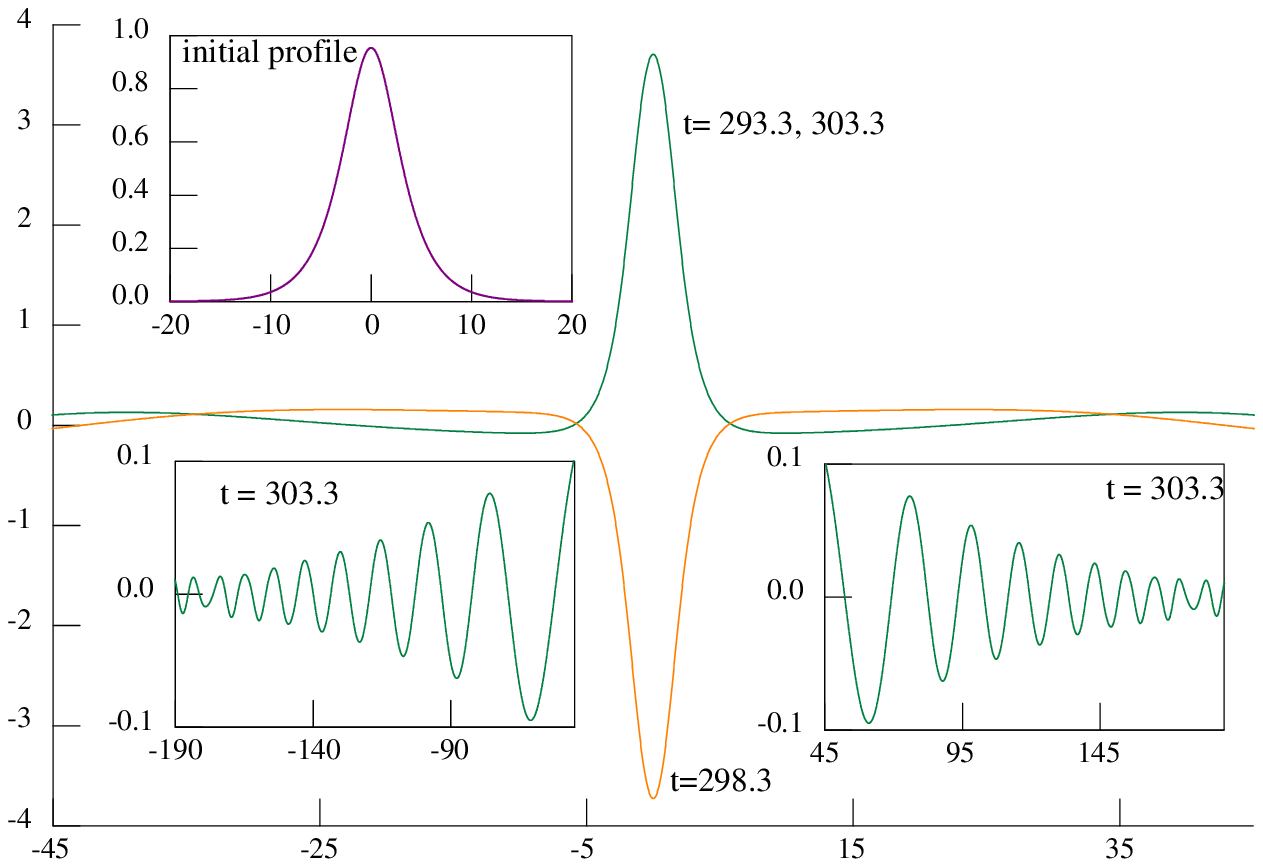}
\caption{\sl Long time behavior of an initially distorted breather: 
$m_1=0,m_2=(0.6,0.7),\beta=0.2,\alpha(t=0)=0.858,~\alpha (t\ra\infty ) =0.314$}
\label{fig3}
\end{figure}

Figure 3 shows a typical case for the breather, and figure 4 one for the wobble.
After a transient that
depends on the magnitude of the distortion, both distorted breathers
and wobbles eventually settle down at a nearby, lower energy,  
stable breather or wobble. 
The excess energy is emitted by means of a trail of phonons both 
in the forward and backward directions, as seen in the insets of figures 3 
and 4.
The trail of phonons resembles an Airy function. 
The pictures in figures 3 and 4 repeat themselves
for longer and longer times.
The relaxation time was found to be of the order of 
$\ds \tau=\frac{1}{|\alpha(t=0)-\alpha(t\ra\infty )|}$.
For $\ds t<\tau$, the parameters
of the wobble and the breather change with time. 
At around $t=\tau$, $\alpha$ and $\beta$ obey again the unitarity
constraint and the wobble and breather do not decay appreciably any more. 
It appears that $\alpha\ra\infty$ depends only on the breather dynamics.
 
\begin{figure}
\epsffile{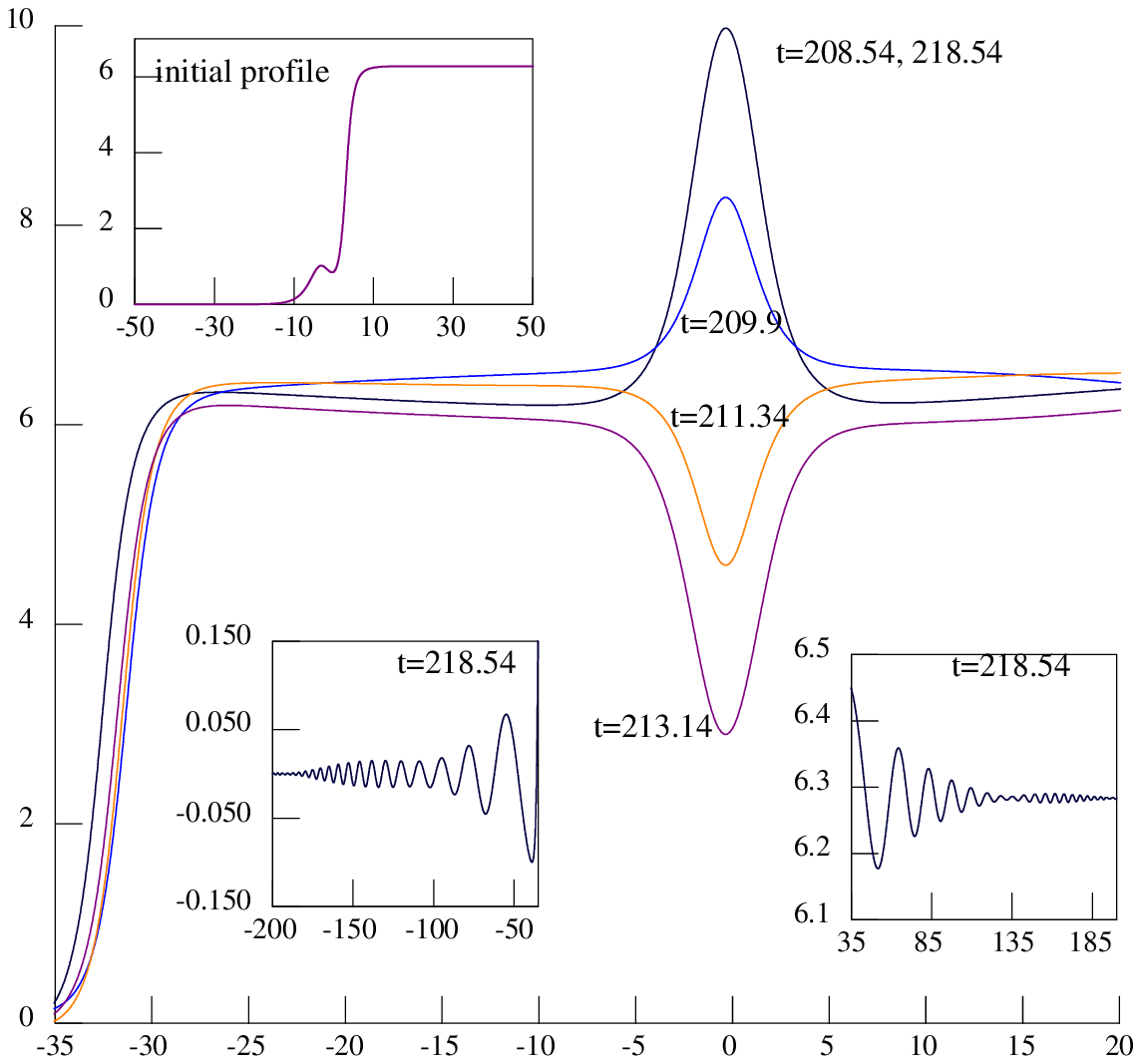}
\caption{\sl Long time behavior of an initially distorted wobble
: $m_1=1, m_2=(0.6,0.7),\beta=0.2,\alpha(t=0)=0.858,
~\alpha (t\ra\infty )= 0.314$}
\label{fig4}
\end{figure}

\section{\sl Distorted kinks and the Wobble}

The existence of a shape mode in the {\sl q=1}
sector of the Sine-Gordon equation has been surrounded by controversy.
The results of the previous section show that there are nonperturbative 
oscillatory shape modes in the {\sl q=1} kink sector. 
The wobble angular frequency 
spectrum fills the gap between the zero mode and the phonon spectrum.
The wobble is an exact solution, whereas the phonon spectrum 
results from approximate linearized solutions
of the Sine-Gordon equation around the kink soliton.
The  wobble must play a role in the
decay dynamics of distorted kinks. 
As we will see below, the shape mode found in the literature,
 is an intermediate stage on the way from the distorted kink to a stable wobble.

Rice\ci{rice}, and later Boesch and Willis\ci{boesch} proposed
the existence of shape modes for distorted kinks in the Sine-Gordon
equation. The angular frequency of the oscillations 
of the kink soliton width found by Rice\ci{rice} and Boesch and 
Willis\ci{boesch} lies above the phonon threshold of
$\ds \omega=1$. The shape mode is therefore unstable to decay into phonons.
Recently, Quintero et al.\ci{quintero} have argued that such a shape mode 
does not exist. Quintero et al.\ci{quintero} claim that 
the behavior observed in the work of Boesch and Willis\ci{boesch} is due
to discretization effects in the numerical calculation. 
They base their hypothesis on a work by Kivshar et al.\ci{kivshar}. 
Kivshar et al.\ci{kivshar} predict the birth of shape modes bifurcating 
from the continuum spectrum of phonons that plunge below threshold upon the
application of perturbations. 
For the unperturbed integrable Sine-Gordon 
equation the effect disappears in the continuum, when 
lattice spacing tends to zero. Kivshar et al.\ci{kivshar} 
state that there are no shape modes for integrable equations.
The results of the previous section show that there is
a whole set of nonperturbative shape modes in the Sine-Gordon equation.

In order to connect the wobble soliton to perturbed kinks, we
followed the evolution of a distorted kink 

\bea\label{kink}
u(x,t) = 4~tan^{-1}({e^{\gamma x}}),
\eea
\noindent
with $\ds |\gamma -1|$ the distortion parameter.
$\gamma (t)$ is extracted from the data by comparing 
to eq.(\ref{kink}). The phenomenological function 
\bea\label{fitfun}
\gamma(t)=1+a~e^{-t^b c}~cos(2\alpha t+\phi),
\eea

{\noindent}captures the broad features of $\ds \gamma(t)$.
From eq.(\ref{fitfun}) we obtained the angular frequency of the oscillation,
$\ds \alpha$. We found that $\ds \alpha$ varies with time.
There is a slow drift of $\alpha$ towards lower values.
This is depicted in figure 5. The angular frequency $\ds\alpha$ of 
eq.(\ref{fitfun}) drops from $\ds 2\alpha = 1.048 $ at around t=200 
to $\ds 2\alpha = 1.0015$ later.
Both the amplitude of the oscillation and the frequency diminish gradually.
The results of figure 5 
agree in general with Boesch and Willis\ci{boesch}.
However, contrarily to the predictions
of the collective coordinate method of Rice\ci{rice}, the frequency is not 
constant.

\begin{figure}
\epsffile{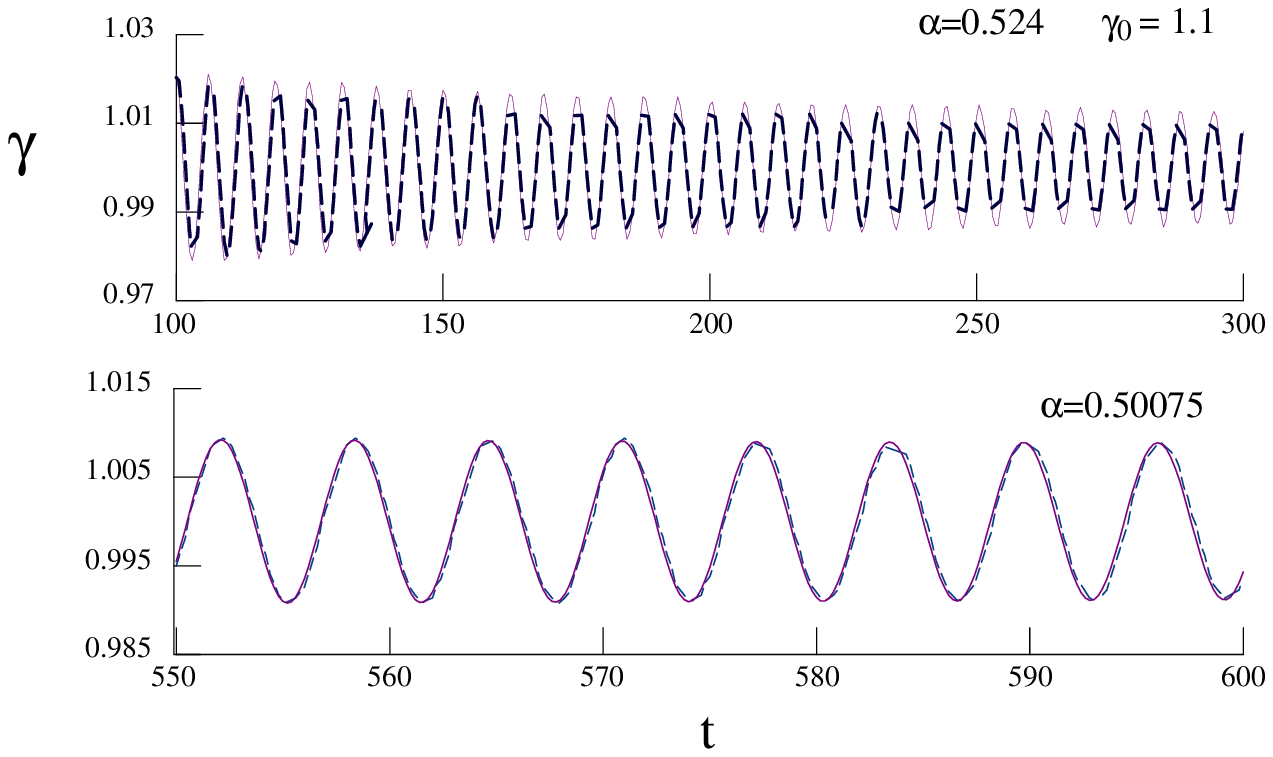}
\caption{\sl Distorted kink width $\gamma$ of eq.(\ref{kink})
as a function of time. Fitted values of eq.(\ref{fitfun}), full curve; input
data, dashed curve}
\label{fig5}
\end{figure}

We consider now the the numerical simulations of Quintero et al.\ci{quintero}.
Quintero et al.\ci{quintero} use an insufficient extent for the x axis 
$\ds L =| x_{max}|= 100$, that does not prevent the reabsorption of reflected 
phonons from the boundary. 
These phonons pump back energy into the oscillating
soliton and blur the picture.
The velocity of propagation of the phonons
is $\ds v =\frac{k}{\sqrt{k^2+1}}$, asymptotically tending towards
$\ds v =1$. Using this asymptotic value, the reflected
phonons collide and feed energy back to the soliton at 
 $\ds t\approx 200$. In figure 2 of \ci{quintero}, and at approximately 
that time, the decaying single frequency shape mode picture starts to 
break down.

Increasing the span of the
integration region with time prevents the reabsorption
of phonons. The slowly drifting single frequency behavior is seen to persist for
longer and longer times. Whether the oscillation frequencies cross the threshold
of $\omega=1$, signaling 
the transition to a stable shape mode lookes unclear from the previous figures.
To accelerate the decay process a very distorted kink is needed. 
Figure 6 shows a case
with $\gamma(t=0)=0.4$, a distortion of 60\% compared to the
unperturbed kink. (The distorted kink energy for this $\gamma$
is $ E = 11.6$ still below
the threshold for the production of a soliton-antisoliton pair at $E =24$).
The angular frequency is now $\ds 2\alpha=0.97$ for
short times. 
A long wavelength modulation of the amplitude is also noticeable
in figure 6.
After t=600, the amplitude of the oscillation appears to stabilize.

As $\alpha$ was obtained by means of a phenomenological function, 
more convincing evidence of the transition to a wobble-like regime
is necessary.

\begin{figure}
\epsffile{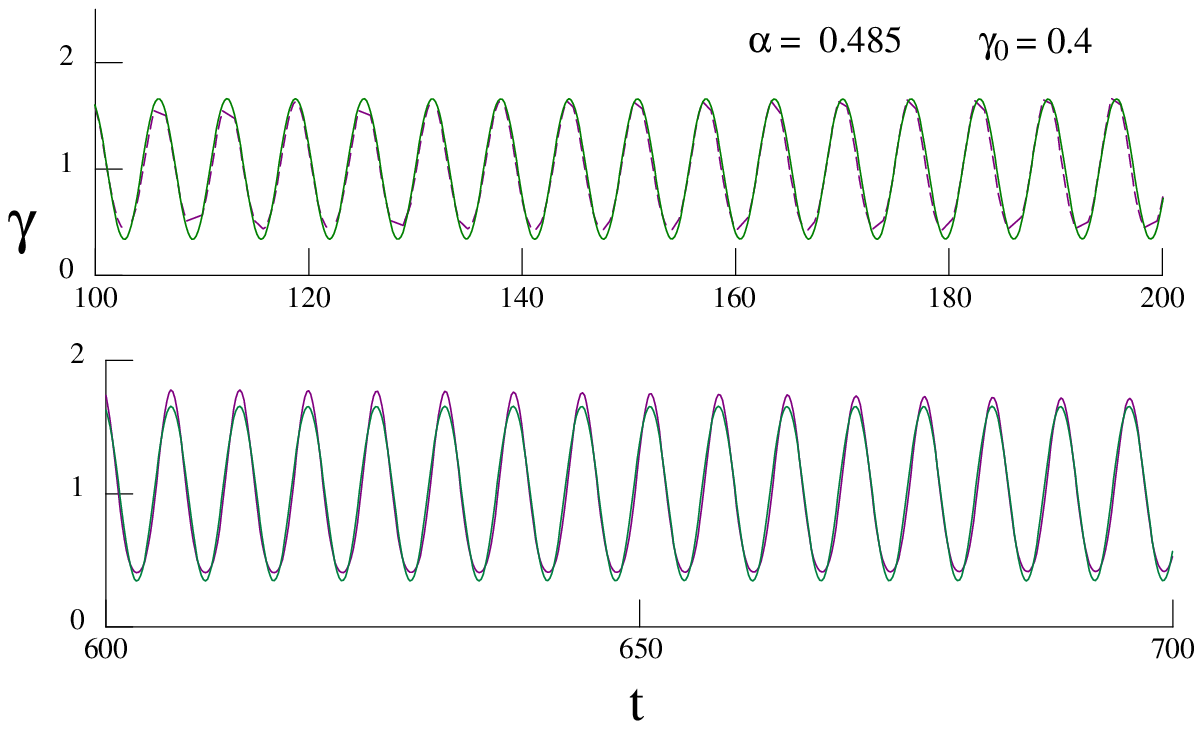}
\caption{\sl Distorted kink width 
$\gamma$ of eq.{kink}
as a function of time, full curve, fitted values, dashed curve, numerical input}
\label{fig6}
\end{figure}
A distorted kink cannot be put in exact correspondence with the
wobble, despite the similarities.
The energy of eq.(\ref{energy}) teaches us that $\ds \beta$ is 
the relevant parameter for the breather admixture to the kink.
Expanding the expression for the wobble of eq.(\ref{wobble}) around 
$\ds \beta = 0$ we find $\ds \gamma (t=0)\approx1-8\beta$.
The factor of 8, and the unitarity constraint that fixes the angular
frequency to be $\ds \alpha=\sqrt{\quart-\beta^2}$ require   
an extremely distorted kink in order to reach a fairly visible 
frequency below threshold. A stronger distortion than $\ds \gamma(t=0)=0.4$ 
as compared to that depicted in figure 6 is necessary.

We therefore considered initial distortions with $\ds \gamma(t=0) < 0.4$.
For such a large intial distortion, the decaying kink profile  
did not match eq.(\ref{kink}) any more. The extraction
of a clean distortion parameter $\ds \gamma(t)$ became impossible. 
The time evolution of the highly distorted kinks leads to
 a completely different object.
Figure 7 shows distorted kinks for $\ds \gamma (t=0)=0.3$ 
at around $ \ds t=290$. 
The profiles resemble very much the wobbles figure 1.

From the graphs one can read off the value of the angular frequency
of the oscillation to be $\ds 2 \alpha\approx 0.92$, well below the phonon
threshold.
The inset shows phonons receding from the center, 
others propagating forwards, are not shown. We performed
long time numerical integrations up to t=1000 and did not see any
sizeable decay of the amplitude of the wobble-like kink.

The kink appears to be decaying to a wobble. If this is indeed the case, 
we should be able to identify the wobble parameters of the decaying 
distorted kink.
\begin{figure}
\epsffile{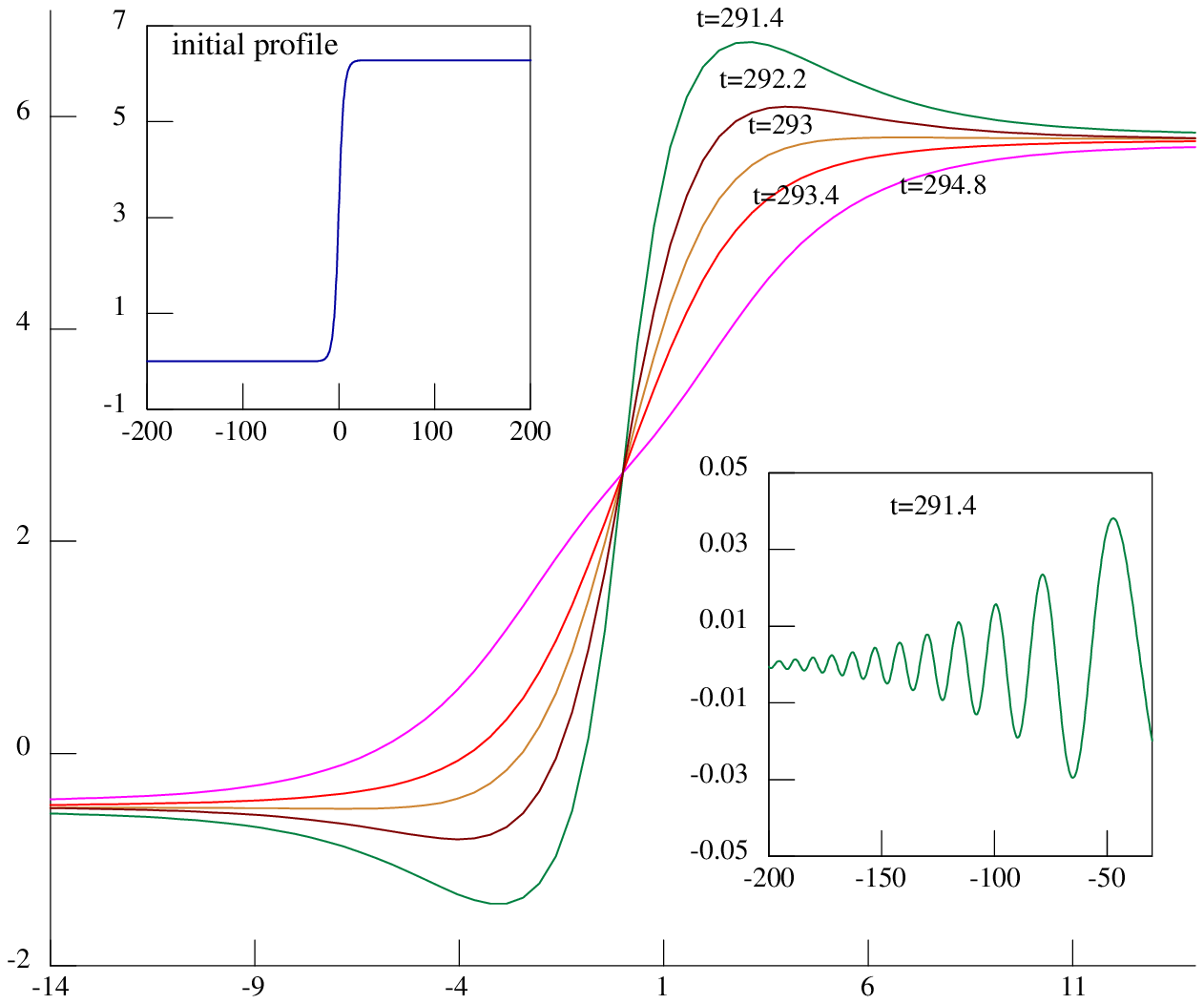}
\caption{\sl Distorted kinks for $\gamma(t=0)=0.3$,
as a function of distance, at various times, showing oscillatory behavior}
\label{fig7}
\end{figure}

To limit the number of free parameters, we first 
considered the $\ds x\ra\infty$ region for both the distorted kink
and the wobble. The wobble has an asymptotic behavior
of $\ds e^{2|\beta |x}$, whereas for the distorted kink
it is $\ds e^{\gamma x}$. If the asymptotic
behavior does not change with time, we have $\ds \gamma \approx 2\beta$.
From this value of $\ds \beta$ and the unitarity constraint, 
$\alpha$ is determined. This $\ds \alpha=\frac{\pi}{T}$ with {\sl T} the
oscillation period, can be readily compared to the numerical results
of figure 7. The agreement is fair, but not satisfactory.
Following the reverse path seemed more appropriate.
The oscillation period of the numerical data 
fixes $\ds \alpha$ and consequently $\ds \beta$ by means of
the unitarity constraint.
There remained three unknown parameters $\ds m_1,~m_{2R},~m_{2I}$ that
were determined using a minimization algorithm.
\begin{figure}
\epsffile{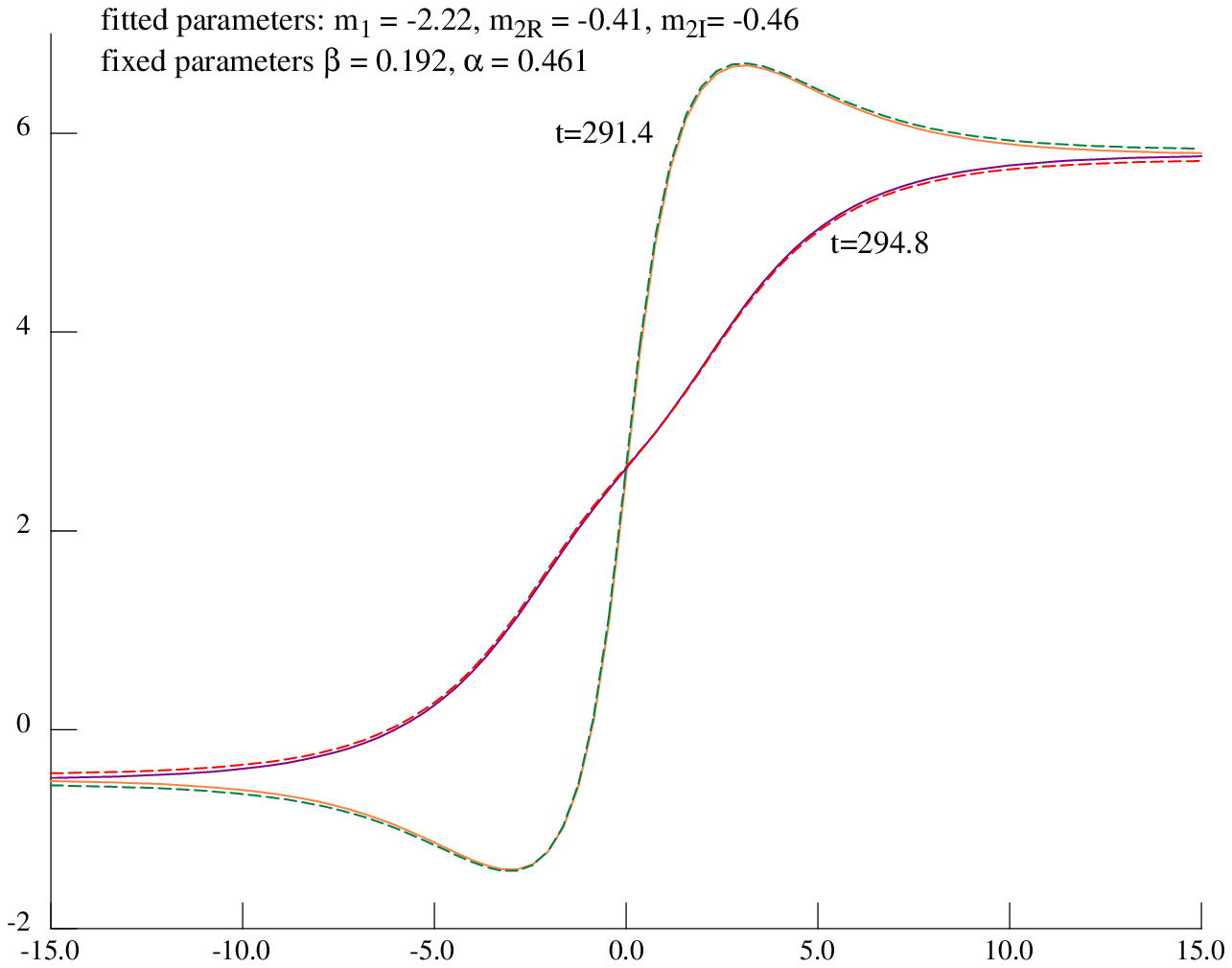}
\caption{\sl Distorted kinks at t=291.4, t=294.8 for $\gamma (t=0)=0.3$,
as a function of distance. Numerical data of figure 7,broken line; 
wobble fit, full line.}
\label{fig8}
\end{figure}

The results are depicted in figure 8.
All the distorted kinks of figure 7 were reproduced with the same parameter set.
The agreement with the data is remarkable, especially so in light 
of the highly nonlinear wobble function.
Long distance discrepancies are due to phonon contributions.

The highly distorted kink has transformed 
into a wobble and a wake of phonons.
The less distorted cases presumably need a much longer time
to reach a wobble. 
For small distortions of the kink, 
it is hard to discern a clear wobble shape.
However, we cannot rule out completely the possibility
of a distorted kink decaying to an undistorted kink and phonons.
In future work we plan to address this and other related problems.

\section{\sl Conclusions}

We have shown that there exists a set of wobbling, apparently stable, 
nonperturbative solutions to the Sine-Gordon equation in the {\sl q=1} sector. 
Highly distorted kinks eventually decay to wobbles and phonons.
The results are relevant to the investigation of scattering events
of Sine-Gordon kinks from impurities.
Other nonlinear equations that support breathers, such as the modified KdV 
equation, may bear wobble solutions too.

The existence of the wobble may have technological implications.
Wobbles produced in Josephson junctions could carry
analogical information with relative stability.

\end{document}